\documentclass[twocolumn,showpacs,preprintnumbers,prl,superscriptaddress,floatfix]{revtex4}
\usepackage{epsfig,amssymb,amsmath,hyperref,bm}
\usepackage{graphics,graphicx,psfrag}
\newcommand\beq{\begin{equation}}
\newcommand\eeq{\end{equation}}
\newcommand\bea{\begin{eqnarray}}
\newcommand\eea{\end{eqnarray}}
\newcommand{\nonum}{\nonumber}

\newcommand\px{\partial_x}

\newcommand\tp{t_{\perp}}
\newcommand\tV{\tilde V}
\newcommand\tU{\tilde U}

\begin{document}

\title{Charge ordering in quasi-one-dimensional systems with frustrating 
interactions}

\author{Siddhartha Lal}
\email{sanjayl@thp.uni-koeln.de}
\affiliation{Institut f\"ur Theoretische Physik, Universit\"at zu K\"oln, 
Z\"ulpicher Stra{\ss}e 77, D-50937 K\"oln, Germany}  
\author{Mukul S. Laad}
\email{M.S.Laad@lboro.ac.uk}
\affiliation{Department of Physics, Loughborough University, LE11 3TU, UK}

%\date{}

\begin{abstract}
Motivated by the co-existing charge and spin order found in 
strongly correlated ladder systems, we study an effective pseudospin 
model on a coupled two-leg ladder. A bosonisation analysis yields a 
rich phase diagram showing Wigner/Peierls charge order
and Neel/dimer spin order. In a broad parameter 
regime, the orbital antiferromagnetic phase is found to be stable. 
An intermediate gapless phase of finite width is found to lie in 
between two charge-ordered gapped phases. Our work is potentially 
relevant for a unified description of a class of strongly correlated, 
quarter-filled ladder systems.
\end{abstract}


%\keywords{}

\maketitle
	
Strongly correlated ladder systems are fascinating candidates 
for studying the interplay of spin and charge ordering, and their 
combined influence on the emergence of novel ground states~\cite{sachdev}. 
Several recent studies provide experimental 
realisations of such systems, exhibiting diverse phenomena
like charge order (CO), antiferromagnetism (AF) and unconventional 
superconductivity (uSC) as functions of suitable control 
parameters~\cite{tokura,yamauchi}.  
The existence of several non-perturbative theoretical techniques
in one dimension have also resulted in the study 
of the emergence of exotic phases from instabilities of the high-$T$ 
Luttinger liquid~\cite{giamarchi}.
Given that these systems are Mott insulators, longer-range 
Coulomb interactions are relevant in understanding CO/AF/uSC phases. 
Inspite of some recent work~\cite{capponi},  
a detailed understanding of the effects of longer-range interactions in 
quasi-1D systems is a largely unexplored problem. 
Further, attention has mostly focussed on studies of models at 1/2-filling
~\cite{giamarchi}. 
\par
Here, we study an effective pseudospin model describing charge degrees 
of freedom of a one-dimensional $1/4$-filled electronic system.  
Such an effective model can be 
derived from an extended Hubbard model (with longer-range interactions)
in a number of physically relevant cases~\cite{fulde,mostovoy,horsch}. It has 
also been suggested~\cite{emery} that this may be a relevant starting 
point for the study of charge-order in organics.  
While the weak coupling limit of the underlying extended Hubbard model
has been recently studied~\cite{donohue,orignac},
we note that the real systems under consideration are generically in 
the strong coupling regime of the model~\cite{fulde,mostovoy,horsch}. 
To the best of our knowledge, this regime has not been studied in sufficient 
detail. 
\par
We start with the effective pseudospin Hamiltonian
\beq
H^{chain} = -\sum_{j} [2t~\tau_{j}^{x} 
+   V~\tau_{j}^{z}\tau_{j+1}^{z}]
\label{onechain}
\eeq
where $2t$ is the transverse field, $V$ is the n.n interaction 
pseudospin coupling.  Here, the ${\bf \tau}$ represent charge 
degrees of freedom in an effective model derived from a more 
basic electronic Hamiltonian for a $1/4$-filled system
~\cite{fulde,mostovoy,horsch}.  
This is the 1D Ising model in a transverse field. 
Rotating the pseudospin axis $\tau^{x}\rightarrow\tau^{z}$, 
$\tau^{z}\rightarrow - \tau^{x}$ and introducing a  
bond-fermion repulsion 
$U_{\perp}\sum_{i,a,b\neq a}(n_{i,a}-n_{i+1,a})(n_{i,b}-n_{i+1,b})$ 
as well as a bond-fermion transfer term 
$t_{\perp}\sum_{i,a,b\neq a}(c_{i,a,\uparrow}^{\dagger}c_{i+1,a,\uparrow}
c_{i,b,\downarrow}^{\dagger}c_{i+1,b,\downarrow} + {\rm h.c})$
between two such systems described by the indices ($a,b$)
~\cite{fulde,mostovoy,horsch}, 
we have the effective Hamiltonian for the charge sector of the 
coupled system in terms of a pseudospin ladder model
\bea
H 
&=& -\sum_{j,a}\lbrack 2t~\tau_{j,a}^{z} 
		+ V~\tau_{j,a}^{x}\tau_{j+1,a}^{x}\rbrack\nonum\\
&& -\hspace*{-0.2cm}\sum_{j,a,b\neq a} \lbrack 
U_{\perp}\tau_{j,a}^{z}\tau_{j,b}^{z} 
+ ~\tp~ (\tau_{j,a}^{x}\tau_{j,b}^{x} 
+ \tau_{j,a}^{y}\tau_{j,b}^{y})\rbrack~, 
\label{twochain}
\eea
where $a,b=1,2$ is the chain index. Having explored the case of 
$U_{\perp} >> V$ in an earlier work ~\cite{first}, we will explore 
the scenario of $U_{\perp} << V$ below. 
In the spin sector, coupling between the two systems leads to a $S=1/2$ 
Heisenberg ladder-type model, and has been studied extensively by 
several authors ~\cite{giamarchi,tsvelik}. Interchain spin coupling turns out 
to be relevant, opening up a spin gap concomitantly with generating 
AF/dimer ordered ground states with spontaneously broken 
SU(2)/translation symmetries.
\par
We now analyse the charge sector. To begin, we introduce 
(bond-)fermion operators via a Jordan-Wigner transformation.
In terms of these fermions, upon denoting the chains as 
$a=\uparrow,\downarrow$, we find an effective Hamiltonian 
for the 1D Hubbard model with an equal-spin pairing term and an 
on-site spin-flip term
\bea
H &=& -\frac{{\tilde t}}{2}\sum_{j,a} (\psi_{j,a}^{\dagger}\psi_{j+1,a}
+ {\rm h.c})
-\frac{\tp}{2}\sum_{j}(\psi_{j,a}^{\dagger}\psi_{j,b} 
+ {\rm h.c})\nonum\\ 
&&\hspace*{-1.1cm} + \tU\sum_{j} n_{j,\uparrow} n_{j,\downarrow} 
+ \tV\sum_{j,a} (\psi_{j,a}^{\dagger}\psi_{j+1,a}^{\dagger} 
+ {\rm h.c}) + \mu\sum_{j,a} n_{j,a}
\label{effham}
\eea
where ${\tilde t}$ and $\tp$ are the in-chain and inter-chain 
hopping parameters respectively, $\mu=-2t$ the chemical 
potential, $\tU = - U_{\perp}$ the on-site (Hubbard) interaction 
coupling and $\tV = V/4$ the pairing strength. 
Note that while we treat the parameters ${\tilde t}$ and $\tV$ as 
independent parameters for the sake of generality, ${\tilde t}=\tV=V$ 
in our original model (\ref{twochain}).
Bosonising in the usual way (in terms of the 
usual charge $\rho = \uparrow + \downarrow$ and spin 
$\sigma = \uparrow - \downarrow $ variables), we obtain the 
effective low-energy bosonic Hamiltonian  
\bea
H &=& \frac{1}{2\pi}\int dx 
[v_{\rho} K_{\rho} (\pi \Pi_{\rho}(x))^{2} + \frac{v_{\rho}}{K_{\rho}} 
(\px \phi_{\rho} (x))^{2}]\nonum\\
&&\hspace*{-0.5cm} + \frac{1}{2\pi}\int dx
 [v_{\sigma} K_{\sigma} (\pi \Pi_{\sigma}(x))^{2} 
+ \frac{v_{\sigma}}{K_{\sigma}} (\px \phi_{\sigma} (x))^{2}]\nonum\\
&&\hspace*{-0.5cm} + \frac{U_{\sigma}}{2\pi\alpha}\int dx 
\cos(\sqrt{8}\phi_{\sigma}) 
+ \frac{U_{\rho}}{2\pi\alpha}\int dx \cos(\sqrt{8}\phi_{\rho})\nonum\\
&&\hspace*{-1.3cm} + \frac{\tV}{2\pi\alpha}\int dx 
\cos(\sqrt{2}\theta_{\rho})\cos(\sqrt{2}\theta_{\sigma}) 
- \frac{\sqrt{2}\mu}{\pi}\int dx \px\phi_{\rho} (x)\nonum\\
&&\hspace*{-1.3cm} + \frac{\tp}{\pi\alpha}\int dx \cos(\sqrt{2}\phi_{\sigma})
\cos(\sqrt{2}\theta_{\sigma}) + \frac{V_{1}}{\pi\alpha}\int dx 
\cos(\sqrt{8}\theta_{\sigma})\nonum\\
&&\hspace*{-1.6cm} + \frac{V_{2}}{\pi\alpha}\int dx \cos(\sqrt{2}\phi_{\sigma})
\cos(\sqrt{2}\theta_{\rho}) + \frac{V_{3}}{\pi\alpha}\int dx 
\cos(\sqrt{8}\theta_{\rho})
\eea  
where 
$\Pi_{\rho} = \frac{1}{\pi} \px\theta_{\rho}$~,~
$\Pi_{\sigma} = \frac{1}{\pi} \px\theta_{\sigma}$~,~
$v_{\rho} K_{\rho} = v_{F} = v_{\sigma} K_{\sigma}$~,~
$v_{\rho}/K_{\rho} = v_{F} (1 + \frac{\tU}{\pi v_{F}})$ and
$v_{\sigma}/K_{\sigma} = v_{F} (1 - \frac{\tU}{\pi v_{F}})$.  
Among the various cosine potentials, 
we have the usual spin-flip backscattering $\cos(\sqrt{8}\phi_{\sigma})$ and 
Umklapp $\cos(\sqrt{8}\phi_{\rho})$ terms as well as the triplet 
superconducting $\cos(\sqrt{2}\theta_{\rho})\cos(\sqrt{2}\theta_{\sigma})$ 
term. The chemical potential term can be absorbed 
by performing the shift $\phi_{\rho}\rightarrow \phi_{\rho} 
+ \frac{\sqrt{2}K_{\rho}\mu}{v_{\rho}} x$.
The cosine potentials with couplings $V_{1}$, $V_{2}$ and $V_{3}$ are 
generated under RG by the $\tp$ and $\tV$ terms. 
We find the RG equations for the various couplings to second-order as 
\bea
\frac{d U_{\rho}}{dl} &=& (2 - 2K_{\rho}) U_{\rho}\nonum\\
\frac{d U_{\sigma}}{dl} &=& (2 - 2K_{\sigma}) U_{\sigma} 
- (\frac{1}{K_{\sigma}} - K_{\sigma})\tp^{2}\nonum\\
\frac{d \tV}{dl} &=& (2 - \frac{1}{2}(\frac{1}{K_{\sigma}} 
+ \frac{1}{K_{\rho}})) \tV - K_{\sigma}\tp V_{2} \nonum\\
\frac{d\tp}{dl} &=&\hspace*{-0.2cm} (2 - \frac{1}{2}(K_{\sigma} 
+ \frac{1}{K_{\sigma}})) \tp \hspace*{-0.1cm}-\hspace*{-0.1cm}
\frac{\tV V_{2}}{K_{\rho}} \hspace*{-0.1cm}-\hspace*{-0.1cm}
(K_{\sigma} U_{\sigma}+\frac{V_{1}}{K_{\sigma}})2\tp\nonum\\
\frac{d V_{1}}{dl} &=& (2 - \frac{2}{K_{\sigma}}) V_{1} 
+ (\frac{1}{K_{\sigma}} - K_{\sigma})\tp^{2}\nonum\\
\frac{d V_{2}}{dl} &=& (2 - \frac{1}{2}(K_{\sigma} 
+ \frac{1}{K_{\rho}})) V_{2} - \frac{\tp\tV}{K_{\sigma}}\nonum\\
\frac{d V_{3}}{dl} &=& (2 - \frac{2}{K_{\rho}}) V_{3}~.
\label{rgeq1}
\eea
The RG equations for the two interaction parameters $(K_{\rho}, K_{\sigma})$ 
as well as the parameter $\delta=K_{\rho}\mu/v_{\rho}$ are found to be 
\bea
\frac{d K_{\sigma}}{dl} &=& 
-K_{\sigma}^{2} U_{\sigma}^{2} + V_{1}^{2}\nonum\\
\frac{d K_{\rho}}{dl} &=& -K_{\rho}^{2} U_{\rho}^{2}
J_{0}(\delta (l)\alpha) \nonum\\
\frac{d\delta}{dl}&=& \delta (l) - U_{\rho}^{2}J_{1} (\delta (l)\alpha)~,
\label{rgeq2}
\eea
where $\delta (l) = \delta e^{l}$, 
$\alpha$ is a short-distance cut-off like the lattice spacing and 
$J_{0}(x), J_{1}(x)$ are Bessel functions~\cite{giamarchi}.
\par
For repulsive interactions ($U_{\perp}>0$) between the bond-fermions, 
the $\sigma$ sector is massless and $K_{\sigma}$ flows 
under RG to the fixed point value $K_{\sigma}^{*} \gtrsim 1$ 
and $1/2\leq K_{\rho}\leq 1$. At 1/2-filling (for the bond-fermions), 
the couplings $U_{\rho}$, 
$\tV$, $\tp$, $V_{1}$ and $V_{2}$ are all relevant while $U_{\sigma}$ 
and $V_{3}$ are irrelevant. 
The competition to reach strong-coupling first is, however, mainly 
between $U_{\rho}$, $\tp$ and $\tV$.
We show below the phase diagram as derived from this analysis.
\begin{figure}[htb]
\begin{center}
\scalebox{1}{
\psfrag{1}[bl][bl][0.9][0]{$K_{\rho}>(1/K_{\sigma},1/4(K_{\sigma}+1/K_{\sigma}))$}
\psfrag{2}[bl][bl][0.9][0]{$1/K_{\sigma}>K_{\rho}>1/4(1/K_{\sigma} + /K_{\rho})$}
\psfrag{3}[bl][bl][0.9][0]{$K_{\sigma}<(1/4(1/K_{\rho}+1/K_{\sigma}),1/4(K_{\sigma} + 1/K_{\sigma}))$}
\includegraphics{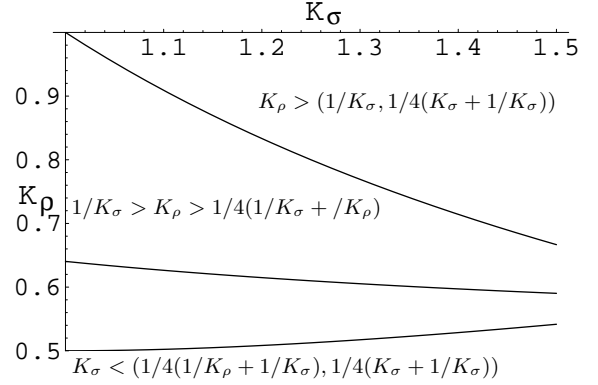}
}
\end{center}
\caption{The RG phase diagram in the $(K_{\sigma}, K_{\rho})$ plane
for repulsive interchain interactions ($U_{\perp}<0$). 
The three regions $K_{\rho}<(1/4(1/K_{\sigma} + 1/K_{\rho}), 
1/4(K_{\sigma} + 1/K_{\sigma}))$, 
$1/K_{\sigma}>K_{\rho}>1/4(1/K_{\sigma} + 1/K_{\rho})$ and 
$K_{\rho}>(1/K_{\sigma}, 1/4(K_{\sigma} + 1/K_{\sigma}))$ give the 
values of $(K_{\sigma},K_{\rho})$ for which the 
couplings $U_{\rho}$, $\tp$ and $\tV$ respectively are the fastest 
to grow under RG.}  
\label{phased1}
\end{figure}
\par
In the phase diagram in Fig.(\ref{phased1}), the three lines with intercepts 
at $(K_{\sigma}=1, K_{\rho}=1)$, $(K_{\sigma}=1, K_{\rho}=0.64)$ and 
$(K_{\sigma}=1, K_{\rho}=1/2)$ are the relations $K_{\sigma}=1/K_{\rho}$, 
$K_{\rho}=1/4(1/K_{\sigma} + 1/K_{\rho})$ and 
$K_{\rho}=1/4(K_{\sigma} + 1/K_{\sigma})$ respectively. The regions 
$K_{\rho}<(1/4(1/K_{\sigma} + 1/K_{\rho}), 1/4(K_{\sigma} + 1/K_{\sigma}))$, 
$1/K_{\sigma}>K_{\rho}>1/4(1/K_{\sigma} + 1/K_{\rho})$ and 
$K_{\rho}>(1/K_{\sigma}, 1/4(K_{\sigma} + 1/K_{\sigma}))$ signify 
the values of $K_{\rho}$ and $K_{\sigma}$ for which $U_{\rho}$ (in-chain 
Wigner charge-ordered Mott insulator), $\tp$ and $\tV$ (in-chain 
Peierls charge-ordered Mott insulator of preformed bond-fermion pairs) 
respectively are the fastest to reach strong-coupling. The RG equations 
for the coupling $U_{\rho}$, the interaction parameter $K_{\rho}$ and 
the incommensuration parameter $\delta$ are familiar from the literature 
on commensurate-incommensurate transitions \cite{japaridze}. For 
temperatures $T>>K_{\rho}\mu$, the finite chemical potential is unable 
to quench the Umklapp scattering processes, allowing for the growth of 
$U_{\rho}$ to strong-coupling. For $T<<K_{\rho}\mu$, the finite chemical 
potential cuts off the RG flow of $U_{\rho}$, freezing the Umklapp 
scattering processes. 
\par
For the case of $\tp$ being the most relevant coupling, 
we find our RG equations to be a non-trivial 
generalisation of those derived in \cite{kusmartsev} for the model of 
two coupled spinless fermion chains without the intrachain $\tV$ 
pairing term. The resulting picture then 
describes strong interchain two-particle correlations between bond-fermions 
sharing a rung. Following the analysis 
outlined in \cite{kusmartsev,tsvelik}, we conclude that this phase is 
a novel insulating phase characterised by a mass gap and delocalised 
bond-fermions on rungs, resembling the orbital antiferromagnetic ground 
state found in the spinless two-chain problem. This matches our finding of 
an orbital antiferromagnetic ground state in the strongly-coupled ladder 
with dominant antiferromagnetic rung-couplings in an earlier 
work~\cite{first}.  
Away from 1/2-filling (for the bond-fermions), the competition is mainly 
between $\tp$ and $\tV$. For $\tV$ reaching strong-coupling ahead 
of $\tp$, the system is in a channel triplet-spin singlet superconducting 
phase with mobile intra-chain hole pairs; for $\tp$ reaching 
strong-coupling ahead of $\tV$, we are currently unable to describe in 
more detail the dominant instability away from the 
orbital antiferromagnetism like insulating phase away from quarter-filling.  
\par
\begin{figure}[htb]
\begin{center}
\scalebox{0.4}{
\psfrag{1}[bl][bl][3][0]{$V_{1}$}
\psfrag{2}[bl][bl][3][0]{$\tp$}
\psfrag{3}[bl][bl][3][0]{I}
\psfrag{4}[bl][bl][3][0]{II}
\psfrag{5}[bl][bl][3][0]{$(0,0)$}
\includegraphics{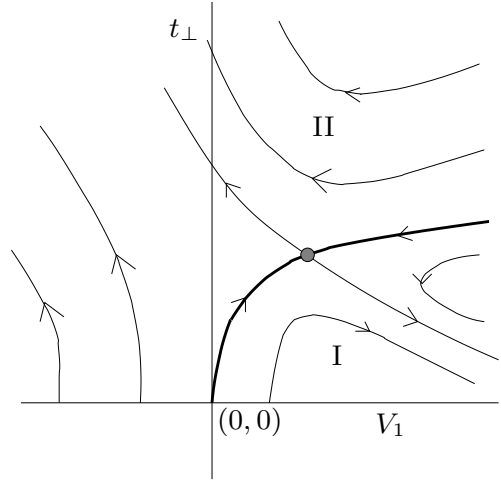}
}
\end{center}
\caption{The RG phase diagram in the $(V_{1}, \tp)$ plane. The thick line 
characterises the set of points which flow to the intermediate fixed point 
$(V_{1}^{*}, \tp^{*})$ shown by the filled circle. The thin lines show 
all RG flows which flow towards strong-coupling in the two phases {\bf I} 
and {\bf II}, characterised by the relevant couplings $\tV$ and $\tp$ 
respectively.} 
\label{intfix}
\end{figure}

Interestingly, while studies of ladder models 
have shown a plethora of charge and spin-ordered gapped phases
~\cite{giamarchi,donohue,orignac}, there remains 
the intriguing possibility that some of these gapped phases may 
themselves be separated from one another by non-trivial gapless phases 
of finite width \cite{tsvelik}. In what follows, we provide an 
explicit realisation of this scenario. 
The RG equations (\ref{rgeq1})-(\ref{rgeq2}) reveal the existence of 
a non-trivial fixed point (FP) for any value of $(K_{\rho},K_{\sigma})$ 
lying the in ranges $1/2<K_{\rho}<1$, $1<K_{\sigma}<2+\sqrt{3}$ and which 
is perturbatively accessible from the trivial weak-coupling FP. We note 
that a similar non-trivial fixed point was found in a study of the 
anisotropic Heisenberg spin-1/2 chain in a magnetic field \cite{duttasen}, 
where the authors derived a set of RG equations which were very similar 
to those found in \cite{kusmartsev}. Here, the non-trivial FP is given by
\bea
\tp^{*} &=& \sqrt{ab}~,~V_{1}^{*} = \frac{K_{\sigma}^{*}+1}{2}{\tp^{*}}^{2}~,~
U_{\sigma}^{*} = \frac{V_{1}^{*}}{2K_{\sigma}^{*}}\nonum\\
\tV^{*} &=& \sqrt{aK_{\rho}^{*}(cK_{\sigma}^{*} - (K_{\sigma}^{*}+1)^{2}ab)} 
~,~ V_{2}^{*} = \sqrt{\frac{b}{a}}\frac{\tV^{*}}{K_{\sigma}^{*}}
\eea 
where
$a = 2-(K_{\sigma}^{*} + 1/K_{\rho}^{*})/2$~,~
$b = 2-(1/K_{\sigma}^{*} + 1/K_{\rho}^{*})/2$ and
$c = 2-(K_{\sigma}^{*} + 1/K_{\sigma}^{*})/2$~.
Further, we can safely make the approximation of the renormalisations of 
$K_{\rho}$ and $K_{\sigma}$ being small at this non-trivial FP 
\cite{duttasen}. The system is 
gapless at this non-trivial FP as well as at points which flow to it. 
The trivial FP has 6 unstable directions ($U_{\rho},\tV,\tp,V_{1},V_{2}$ 
and $\delta$), 2 stable directions ($U_{\sigma}$ and $V_{3}$) and 2 marginal 
directions ($K_{\rho}$ and $K_{\sigma}$). The non-trivial FP has 5 unstable 
directions, 3 stable directions and two marginal directions. Apart from 
the stable direction given by the $V_{3}$, the presence of the other two
stable directions at the non-trivial FP indicates the existence of a 
two-dimensional surface of gapless theories in the five-dimensional 
$(U_{\sigma},\tV,\tp,V_{1},V_{2})$ coupling space. This gapless phase is 
the analog of the ``Floating Phase" found in the phase diagram of the 
1D axial next nearest neighbour Ising model \cite{duttasen}. We 
present in Fig.(\ref{intfix}) below a RG flow phase diagram which 
is projected onto the 
$(V_{1}, \tp)$ plane (a similar RG flow diagram is found for the case 
of the anisotropic Heisenberg model in a magnetic field $h$ \cite{duttasen} 
in the $(a,h)$ plane, where $a$ is the anisotropy parameter). 
\par
The regions {\bf I} and {\bf II} characterise all RG flows 
which do not flow to the intermediate fixed point at 
$(V_{1}^{*}, \tp^{*})$. In region {\bf I}, $V_{1}$ flows to 
strong-coupling while $\tp$ decays; for $1/2<K_{\rho}<1$ and 
$K_{\sigma}>1$, we know from the above discussion that in this region, 
the coupling $\tV$ will reach strong-coupling first. In region {\bf II}, 
both $\tp$ and $V_{1}$ grow under RG, with the coupling $\tp$ being the 
first to reach strong-coupling. Thus, the RG trajectory leading to the 
intermediate fixed point represents a gapless phase separating the 
two gapped, charge-ordered phases {\bf I} and {\bf II} characterised 
by the relevant couplings $\tV$ and $\tp$ respectively.
\par
For attractive interactions ($U_{\perp}<0$) between the 
bond-fermions, we can carry out a similar analysis. In this case, 
we can see that $K_{\rho}>1$ while $K_{\sigma}<1$. 
Then, from the RG equations given above, we can see that the Umklapp coupling 
$U_{\rho}$ and $V_{1}$ are irrelevant while the 
couplings $\tp$, $U_{\sigma}$, $\tV$, $V_{2}$ and $V_{3}$ are 
relevant. The competition to reach strong-coupling first is, however, 
mainly between $U_{\sigma}$, $\tp$ and $\tV$. 
We show below the phase diagram at 1/2-filling for the bond-fermions 
as derived from this analysis.   
\begin{figure}[htb]
\begin{center}
\scalebox{1}{
\psfrag{1}[bl][bl][1][0]{$K_{\sigma}>(1/K_{\rho},1/4(1/K_{\rho}+1/K_{\sigma}))$}
\psfrag{2}[bl][bl][1][0]{$1/K_{\rho}>K_{\sigma}>1/\sqrt{3}$}
\psfrag{3}[bl][bl][1][0]{$K_{\sigma}<(1/4(1/K_{\rho}+1/K_{\sigma}),1/\sqrt{3})$}
\includegraphics{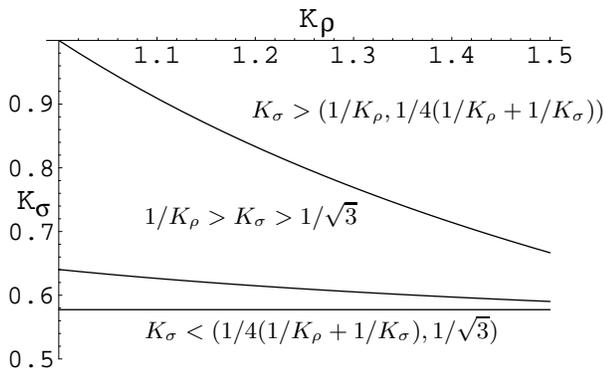}
}
\end{center}
\caption{The RG phase diagram in the $(K_{\rho}, K_{\sigma})$ plane 
for attractive interchain interactions ($U_{\perp}>0$). The three regions 
$K_{\sigma}<(1/4(1/K_{\sigma} + 1/K_{\rho}), 
1/\sqrt{3})$, $1/K_{\rho}>K_{\sigma}>1/\sqrt{3}$ and 
$K_{\sigma}>(1/K_{\rho}, 1/4(1/K_{\rho} + 1/K_{\sigma}))$
give the values of $(K_{\sigma},K_{\rho})$ for which the 
couplings $U_{\sigma}$, $\tp$ and $\tV$
respectively are the fastest to grow under RG.
} 
\label{phased2}
\end{figure}
\par
In the phase diagram in Fig.(\ref{phased2}), the three lines with intercepts 
at $(K_{\rho}=1, K_{\sigma}=1)$, $(K_{\rho}=1, K_{\sigma}=0.64)$ and 
$(K_{\rho}=1, K_{\sigma}=1/\sqrt{3})$ are the relations 
$K_{\sigma}=1/K_{\rho}$, 
$K_{\sigma}=1/4(1/K_{\sigma} + 1/K_{\rho})$ and $K_{\sigma}=1/\sqrt{3}$ 
respectively. The regions $K_{\sigma}<(1/4(1/K_{\sigma} + 1/K_{\rho}), 
1/\sqrt{3})$, $1/K_{\rho}>K_{\sigma}>1/\sqrt{3}$ and 
$K_{\sigma}>(1/K_{\rho}, 1/4(1/K_{\rho} + 1/K_{\sigma}))$ signify 
the values of $K_{\rho}$ and $K_{\sigma}$ for which $U_{\sigma}$ (rung-dimer 
insulator with in-chain Wigner charge-ordering), $\tp$ 
and $\tV$ (insulator with in-chain dimers and Peierls charge-ordering) 
respectively are the fastest to reach strong-coupling.
This matches our finding of a ground state with in-chain Wigner charge 
order and rung-dimers in the strongly-coupled ladder 
with large ferromagnetic rung-couplings in an earlier 
work~\cite{first}. 
Away from 1/2-filling (for the bond-fermions), depending on which of 
the three couplings $\tp$, $\tV$ and $U_{\sigma}$ is the first to reach 
strong-coupling, the system exists either as a superconductor 
with intra-chain hole pairs ($\tV$) or a superconductor with 
rung-singlet hole pairs ($U_{\sigma}$) or a phase reached by following 
the dominant instability away from the orbital antiferromagnetism like 
insulating phase ($\tp$) but which we are currently unable to describe 
in greater detail.
\par
To conclude, we have studied a model of strongly correlated coupled 
quasi-1D systems at $1/4$-filling using an effective 
pseudospin ladder model with 
$V> t,t_{\perp},U_{\perp}$.  Using a bosonisation analysis, we 
find two different types
of charge/spin ordered ground states at $1/4$-filling. Transverse 
bond-fermion hopping is found to stabilise a new, gapped (insulating) 
phase characterised by interchain two-particle coherence of a type 
resembling orbital antiferromagnetism~\cite{kusmartsev,tsvelik}. 
The spin fluctuations are described by a $S=1/2$ Heisenberg  
ladder-type model for all cases studied here: the spin excitations are 
always massive. Away from this filling, either intra- or interchain 
superconductivity in a gapped spin background is found to be the stable 
ground state. We also find the existence of an intermediate gapless 
phase lying in between two gapped, charge-ordered phases (characterised 
by the relevant couplings $\tV$ and $\tp$ respectively) in the RG phase 
diagram of our model. Our analysis is especially relevant to ladder 
systems like $Sr_{14-x}Ca_{x}Cu_{24}O_{41}$ and $\alpha-NaV_{2}O_{5}$ or 
$\beta-Na_{0.33}V_{2}O_{5}$ (a superconductor) which exhibit charge/spin 
long range order at $x=0$ and superconductivity beyond 
under pressure and/or doping~\cite{tokura,yamauchi}.

\begin{acknowledgments}
SL and MSL thank the DFG (Germany) and EPSRC (UK) respectively for 
financial support. 
\end{acknowledgments}

\end{document}